\documentclass[aps,prc,superscriptaddress,a4,nofootinbib,preprint,showpacs]{revtex4}

\usepackage[pdftex]{graphicx,color}
\usepackage{natbib,subfigure}
\usepackage{amsmath,amssymb,latexsym}
\usepackage{pgf}
\usepackage{bm}
\usepackage{tabularx}
\usepackage{url}

\usepackage[colorlinks=true,linkcolor=blue,citecolor=blue]
{hyperref}

\allowdisplaybreaks

\newcommand{\mun}{\mu_n}
\newcommand{\mup}{\mu_p}
\newcommand{\mue}{\mu_e}
\newcommand{\mumu}{\mu_{\mu}}
\newcommand{\mul}{\mu_{\Lambda}}

\newcommand{\sllone}{${\rm S}\Lambda\Lambda 1$ }
\newcommand{\slltwo}{${\rm S}\Lambda\Lambda 2$ }
\newcommand{\sllsam}{${\rm S}\Lambda\Lambda 3$ }
\newcommand{\sllonep}{${\rm S}\Lambda\Lambda 1'$ }
\newcommand{\sllsamp}{${\rm S}\Lambda\Lambda 3'$ }

\usepackage{ulem}

\begin{document}
\title{Hyperon Puzzle of Neutron Stars with Skyrme Force Models}


\author{Yeunhwan \surname{Lim}}
\email{ylim9057@ibs.re.kr}
\affiliation{Rare Isotope Science Project,
Institute for Basic Science,
Daejeon 305-811, Republic of Korea}

\author{Chang Ho \surname{Hyun}}
\email{hch@daegu.ac.kr}
\affiliation{Department of Physics Education,
Daegu University, Gyeongsan 712-714, Republic of Korea}

\author{Kyujin \surname{Kwak}}
\email{kkwak@unist.ac.kr }
\affiliation{School of Natural Science,
Ulsan National Institute of Science and Technology (UNIST), Ulsan 689-798, Republic of Korea}

\author{Chang-Hwan \surname{Lee}}
\email{clee@pusan.ac.kr}
\affiliation{Department of Physics,
Pusan National University, Busan 609-735, Republic of Korea}
\date{November 2, 2015}

\begin{abstract}
%
We consider the so called hyperon puzzle of neutron star (NS).
We employ Skyrme force models for the description of in-medium nucleon-nucleon,
nucleon-Lambda hyperon ($N\Lambda$), and Lambda-Lambda ($\Lambda\Lambda$)
interactions. A phenomenological finite-range force for the $\Lambda\Lambda$ 
interaction is considered as well.
Equation of state (EoS) of NS matter is obtained in the framework of density functional
theory, and Tolman-Oppenheimer-Volkoff equations are solved to obtain the mass-radius
relations of NSs.
It has been generally known that the existence of hyperons in the NS matter is not well
supported by the recent discovery of large-mass NSs ($M \simeq 2 M_\odot$) since
hyperons make the EoS softer than the one without them.
For the selected interaction models, $N\Lambda$ interactions reduce the maximum mass
of NS by about 30~\%, while $\Lambda\Lambda$ interactions can give about 10~\%
enhancement.
Consequently, we find that some Skyrme force models predict the maximum mass of NS
consistent with the observation of $2 M_\odot$ NSs, and at the same time 
satisfy observationally constrained mass-radius relations.
\end{abstract}


\maketitle

\section{Introduction}

\def\chlee#1{\marginpar[$\Rightarrow$]{$\Leftarrow$}{\bf \em CHL: #1}}
\def\kwak#1{\marginpar[$\Rightarrow$]{$\Leftarrow$}{\bf \em KJK: #1}}

The existence of a neutron stars have been confirmed observationally, 
and the measured masses and radii of neutron stars have been 
used to partially constrain the physical properties of the dense nuclear matter. 
So far, about 70 NSs have been observed and their distribution in mass has been 
accumulated according to their binary types \cite{lattimer2012}. 
In the X-ray/optical binaries, the mean and error-weighted mean of NS mass distribution 
are $1.568 M_\odot$ and $1.368 M_\odot$, respectively, which implies that the measurement 
errors are relatively large. 
The NS-white dwarf binaries have similar precision in the mass distribution to that 
of the X-ray/optical binaries; the mean and error-weighted mean are 
$1.543 M_\odot$ and $1.369 M_\odot$, respectively. 
On the other hand, the NS-NS binaries show relatively stable statistics; the mean and error-weighted
mean are $1.322 M_\odot$ and $1.402 M_\odot$, respectively.
In all of these binaries, 
the mean masses do not exceed $1.6 M_\odot$,
and one may conclude that most of NS masses lie in the range
of $1.2 M_\odot - 1.6 M_\odot$, 
which can be regarded as a canonical range of NS masses.
Note that NSs with larger masses have relatively larger measurement errors except for
a few exceptions for the large NS masses, which were measured with high accuracy, 
e.g., PSR J1614-2230, $(1.97 \pm 0.04) M_\odot$ \cite{1614}
and PSR J0348+0432, $(2.01\pm 0.04)M_\odot$ \cite{0348}.

By solving the Tolman-Oppenheimer-Volkoff (TOV) equation, 
one can see that the NS with large mass can have  
a high-density region in the core, providing 
a room for the existence of non-nucleonic degrees of freedom such as hyperons, Bose-Einstein
condensation, free quark phase and etc.
These exotic states relieve the pressure exerted by the Fermionic nature of nucleons,
make the equation of state (EoS) soft, and eventually reduce the mass of NS.
Among diverse possibilities, hyperons have been playing a key role in reducing the
maximum mass of NS to the canonical range. 
However, recent observations of NSs with 
$(1.97 \pm 0.04) M_\odot$ and  $(2.01\pm 0.04)M_\odot$ initiate a challenge to the existence 
of hyperons at the core of NS (so called ``hyperon puzzle").

In order to solve the hyperon puzzle, some studies based upon both relativistic
and non-relativistic mean field (RMF and NRMF, respectively) models have been conducted 
so far. 
The modern version of these 
models shows great accuracy in reproducing the properties of known nuclei. However, 
the maximum mass of NS calculated with these models shows drastic fluctuations, ranging
from $1.4 M_\odot$ to $2.8 M_\odot$  even without including any hyperon yet \cite{lattimer2007}.
Main uncertainty comes from the extrapolation of models to the high density
where no experimental data available.
Therefore, the measurement of the 
$\sim 2 M_\odot$ NS from PSR J1614-2230 and PSR J0348+0432 can rule out some of RMF or NRMF 
models that predict the maximum mass less than $2 M_\odot$.

When the effect of hyperons is included in the RMF models, 
the coupling constants of hyperons are fitted to the depth of hyperon-nucleus
optical potentials at the nuclear saturation density in most of the RMF models.
(Note that the hyperon-nucleus optical potentials are obtained 
 from the single-strangeness hypernuclei.) 
With this method, the RMF models show the similar effect of hyperons on NS mass 
regardless of the differences among various RMF models; the mass of NS is depending on the background nuclear models 
\cite{weiss2012}.
In order to solve the hyperon puzzle, some recent works introduced repulsive forces 
caused by the vector-meson exchange in the RMF models  \cite{weiss2013}. 
In Ref. \cite{hyun2007}, by assuming a hard core of hyperons, one of the authors (CHH) 
obtained the maximum mass of NS close to $2 M_\odot$.
However, the parameterization used in these approaches is not directly compatible with hypernuclear
data and relies partly on the theoretical arguments. 
Dealing with the interactions between hyperons is in a worse situation 
because of statistically insufficient data from the 
double-hypernuclei experiments. 
In most cases, the coupling constants between
hyperons are determined from the symmetry arguments. 
The coupling constants obtained in this manner show very minor effects on the mass of NS 
\cite{bed2005, ryu2007, ryu2009, aa2012}, so in many works the interactions between hyperons are 
justified to be omitted for simplicity.

The purpose of this work is to explore the effect of the $\Lambda\Lambda$ interactions on the
EoS and the resulting bulk properties of NS 
by employing a few Skyrme-type models and a finite-range model 
that reproduce the empirical binding energies of the double-$\Lambda$ hypernuclei. 
For the nucleon and single-$\Lambda$ interactions and the nucleon-nucleon interactions, 
we also use the Skyrme-type models that are consistent with the experimental data. 
We find that our results with these models turn out to be 
quite different from those of the RMF models, suggesting a possible solution 
to the hyperon puzzle. We find that some of our models both predict 
the maximum mass of $2 M_\odot$ and are in good agreement with 
the observationally-constrained mass-radius relation \cite{steiner2010}. 

We organize this paper as follows. Sect. II explains the Skyrme-type models for the 
nucleon-nucleon ($NN$) interactions and the nucleon-$\Lambda$ ($N\Lambda$) interactions. 
As mentioned above, we use both the Skyrme-type models and a finite-range force model 
for the $\Lambda\Lambda$ interactions. Sect. II also contains some details of these 
models. Sect. III is devoted to the results about the properties of NS matter,
and the consequent mass-radius relation of NS.
We summarize our work and give some discussions in Sect. IV.


\section{Interactions}

We employ the standard Skyrme-type forces for the $NN$ interactions given in the form
\begin{eqnarray}
v_{NN}(\bm{r}_{ij}) &=& t_0 (1+x_0 P_\sigma) \delta(\bm{r}_{ij}) +
\frac{1}{2} t_1 (1+x_1 P_\sigma) [ \bm{k}'^2_{ij} \delta(\bm{r}_{ij})
+ \delta(\bm{r}_{ij}) \bm{k}^2_{ij}] \nonumber \\
&& + t_2(1+x_2 P_\sigma) \bm{k}'_{ij} \cdot \delta(\bm{r}_{ij}) \bm{k}_{ij}
+ \frac{1}{6} t_3(1+x_3 P_\sigma) \rho_N^\epsilon (\bm{R}) \delta(\bm{r}_{ij}) \nonumber \\
&& 
+ i W_0 \bm{k}'_{ij} \cdot \delta(\bm{r}_{ij}) (\bm{\sigma}_i + \bm{\sigma}_j) 
\times \bm{k}_{ij},
\end{eqnarray}
where $\bm{r}_{ij} = \bm{r}_i - \bm{r}_j$, $\bm{R} = (\bm{r}_i + \bm{r}_j)/2$,
$\bm{k}_{ij} = -i (\overrightarrow{\nabla}_i - \overrightarrow{\nabla}_j)/2$, 
$\bm{k}'_{ij} = i (\overleftarrow{\nabla}_i - \overleftarrow{\nabla}_j)/2$, 
the spin exchange operator
$P_\sigma = (1+\bm{\sigma}_i \cdot\bm{\sigma}_j)/2$, and the nucleon
density $\rho_N = \rho_n + \rho_p$.
We choose three models SLy4, SkI4, and SGI for the parameters of $NN$ interactions.
As shown in Ref.~\cite{lim2014}, these models are 
in reasonable agreement with the mass-radius constraints of the observed NS 
presented in Ref.~\cite{steiner2010}. 
The basic properties of the models are summarized in Table \ref{tab:nnforce}.
We note that the models also satisfy the maximum mass constraints given by
PSR J1614-2230 or PSR J0348+0432.
\begin{table}
\begin{center}
\begin{tabular}{lccccccc}
\hline
$NN$ model & ~~~~$\rho_0$ ~~~~ & ~~~ $B$ ~~~ & 
~~~ $S_v$ ~~~ & ~~~ $L$ ~~~ & ~~~ $K$ ~~~ & 
~$m^*_N/m_N$ ~ & $M_{\rm max}/M_\odot$ \\ \hline
SLy4 & 0.160 & 16.0 & 32.0 & 45.9 & 230 & 0.694 & 2.07 \\ 
SkI4 & 0.160 & 16.0 & 29.5 & 60.4 & 248 & 0.649 & 2.19 \\ 
SGI  & 0.155 & 15.9 & 28.3 & 63.9 & 262 & 0.608 & 2.25 \\ 
 \hline
\end{tabular}
\end{center}
\caption{Nuclear matter properties and the maximum mass of NS
calculated from the $NN$ Skyrme force models. 
$\rho_0$: saturation density in unit of fm$^{-3}$, 
$B$: binding energy of the symmetric nuclear matter in unit of MeV, 
$S_v$: symmetry energy at the saturation density in unit of MeV, 
$L$: slope of the symmetry energy at the saturation density in unit of MeV, 
$K$: compression modulus of the symmetric matter at the
saturation density in unit of MeV,
$m^*_N/m_N$: ratio of the effective mass of the 
nucleon at the saturation density ($m^*_N$) to the free mass of the nucleon ($m_N$), 
and $M_{\rm max}/M_\odot$: maximum mass of NS in unit of the solar 
mass ($M_\odot$).}
\label{tab:nnforce}
\end{table}

The Skyrme-type interaction of a $\Lambda$ hyperon in the nuclear medium 
was first proposed in Ref.~\cite{rayet1981}, 
and later a two-body density-dependent $N\Lambda$ term replaced the
three-body $N\Lambda\Lambda$ interactions in Ref.~\cite{lans1997}.
The potential for the $N\Lambda$ interaction in our work takes the form
\begin{eqnarray}
v_{N\Lambda} &=& u_0 (1+y_0 P_\sigma) \delta(\bm{r}_{N\Lambda}) 
+\frac{1}{2} u_1 \left[ \bm{k}'^2_{N\Lambda} \delta(\bm{r}_{N\Lambda})
+ \delta(\bm{r}_{N\Lambda})  \bm{k}^2_{N\Lambda}\right] \nonumber \\
&& + u_2 \bm{k}'_{N\Lambda} \cdot \delta(\bm{r}_{N\Lambda})\bm{k}_{N\Lambda} 
+ \frac{3}{8} u'_3 (1+y_3 P_\sigma) \rho^\gamma_N\left(\frac{\bm{r}_N+\bm{r}_\Lambda}{2}\right) \delta(\bm{r}_{N\Lambda}),
\end{eqnarray}
where definitions of $\bm{r}_{N\Lambda}$ and $\bm{k}_{N\Lambda}$ follow
the same convention as in the $NN$ interaction.
There are several parameter sets for the coupling constants in the $N\Lambda$ potential
\cite{ybz1988,fer1989,lans1997,gul2012}. 
We adopt the parameters from an old model, YBZ6 in Ref.~\cite{ybz1988} 
and the most recent ones, HP$\Lambda 2$ and O$\Lambda2$ in Ref.~\cite{gul2012}.
Table \ref{tab:nlam} shows the parameters we use for the $N\Lambda$ interactions.

\begin{table}
\begin{center}
\begin{tabular}{ccccccccc}
\hline
$N\Lambda$ model & ~~~$\gamma$ ~~~& ~~~$u_0$~~~ & ~~~$u_1$~~~
 &  ~~~$u_2$~~~ &  ~~~$u_3^{\prime}$~~~ & ~~~$y_0$~~~ & ~~~$y_3$~~~ & $U^{\rm opt}_\Lambda$ \\
\hline
HP$\Lambda 2$ & $1$   & $-399.946$ & $83.426$& $11.455$& $2046.818$ & $-0.486$ & $-0.660$ & $-31.23$ \\
O$\Lambda 2$ & $1/3$   & $-417.7593$ & $1.5460$& $-3.2671$& $1102.2221$ & $-0.3854$ & $-0.5645$ & $-28.27$ \\
YBZ6 & $1$ & $-372.2$ & $100.4$& $79.60$ &$2000.$ & $-0.107$& $0.$ & $-29.73$ \\
\hline
\end{tabular}
\end{center}
\caption{Parameters for the $N\Lambda$ interactions.
$u_0$ is in unit of MeV$\cdot$ fm$^{3}$, $u_1$ and $u_2$ in unit of MeV$\cdot$ fm$^{5}$, 
and $u'_3$ in unit of MeV$\cdot$ fm$^{3+3\gamma}$. $y_0$ and $y_3$ are dimensionless.
The last column $U^{\rm opt}_\Lambda$ is the depth of the $\Lambda$-nucleus optical potential 
in unit of MeV at the saturation density.}
\label{tab:nlam}
\end{table}

For the interactions between $\Lambda$ hyperons, a Skyrme-type force in the standard form was
proposed in Ref.~\cite{lans1998}, where three sets of model parameters,
\sllone, \slltwo, and \sllsam 
were obtained
from the fit to the binding energy $B_{\Lambda\Lambda}$ of a double-$\Lambda$ hypernucleus 
$^{13}_{\Lambda\Lambda}{\rm B}$.
Recently new parameters have been obtained by considering the fission barrier of 
the actinide nuclei with two $\Lambda$ hyperons \cite{minato2011}, and the models 
we take into account in this work are labeled as \sllonep 
and \sllsamp.
The Skyrme-type force is written as
\begin{eqnarray}
v_{\Lambda\Lambda}(\bm{r}_{ij}) &=&  \lambda_0 \delta(\bm{r}_{ij})
+ \frac{1}{2} \lambda_1 \left[ \bm{k}'^2_{ij} \delta(\bm{r}_{ij})
+ \delta(\bm{r}_{ij}) \bm{k}^2_{ij}\right] 
+ \lambda_2 \bm{k}'_{ij} \cdot \delta(\bm{r}_{ij}) \bm{k}_{ij} 
+ \lambda_3 \rho^\alpha_N (\bm{R}) \delta(\bm{r}_{ij}),
\end{eqnarray}
and the model parameters are shown in Table  \ref{tab:lamlam}.
\begin{table}
\begin{center}
\begin{tabular}{cccccc}
\hline 
$\Lambda\Lambda$ model &  ~~~ $\lambda_0$ ~~~  
 &  ~~~ $\lambda_1$  ~~~ &  ~~~ $\lambda_2$ ~~~ 
  & ~~~ $\lambda_3$ ~~~ &  ~~~ $\alpha$ ~~~\\
\hline
S$\Lambda\Lambda$1& $-312.6$   & $57.5$ & $0$   & $0$& $-$\\
S$\Lambda\Lambda$2& $-437.7$   & $240.7$ & $0$ & $0$& $-$ \\
S$\Lambda\Lambda$3& $-831.8$ & $922.9$ & $0$   & $0$& $-$ \\
\sllonep & $-37.9$ & $14.1$ & 0 & 0 & $-$ \\
\sllsamp & $-156.4$ & $347.2$ & 0 & 0 & $-$ \\
\hline
\end{tabular}
\end{center}
\caption{Parameters of the $\Lambda\Lambda$ interactions in the Skyrme-type force.
$\lambda_0$ is in unit of ${\rm MeV}\cdot {\rm fm}^3$, and $\lambda_1$ in unit of
${\rm MeV}\cdot {\rm fm}^5$. Notice that all models considered here do not have $\lambda_2$ momentum interaction and $\lambda_3$, $\alpha$
density dependent interaction.}
\label{tab:lamlam}
\end{table}
Table \ref{tab:lambinding} shows the $\Lambda\Lambda$ binding energy 
$B_{\Lambda\Lambda}$ calculated with
SLy4, HP$\Lambda2$, and three S$\Lambda\Lambda$ models \cite{gul2014}.
\begin{table}
\begin{center}
\begin{tabular}{ccccc}
\hline
Nuclei & $B_{\Lambda\Lambda}$(\sllone) & $B_{\Lambda\Lambda}$(\slltwo) & $B_{\Lambda\Lambda}$(\sllsam) & $B_{\Lambda\Lambda}$(Exp.) \\
\hline
$^6_{\Lambda\Lambda}{\rm He}$ & 11.88 & 9.25 & 7.60 & $6.93 \pm 0.16$ \cite{naka2010} \\
$^{10}_{\Lambda\Lambda}{\rm Be}$ & 19.78 & 18.34 & 15.19 & $14.94 \pm 0.13$ \cite{dany1963}
\footnote{The value is obtained by adding 2+ excitation energy 3.04 MeV to the experimentally 
deduced value $11.90 \pm 0.13$ MeV with the assumption that 2+ core excitation energies in
$^9{\rm Be}_{\Lambda\Lambda}$ is equal to that of $^{10} {\rm Be}_{\Lambda\Lambda}$.}\\ 
$^{11}_{\Lambda\Lambda}{\rm Be}$ & 20.55 & 19.26 & 16.27 & $20.49 \pm 1.15$ \cite{naka2010}\\ 
$^{12}_{\Lambda\Lambda}{\rm Be}$ & 21.10 & 19.97 & 17.18 & $22.23 \pm 1.15$ \cite{naka2010}\\
$^{13}_{\Lambda\Lambda}{\rm B}$ & 21.21 & 20.26 & 17.76 & $23.30 \pm 0.70$ \cite{naka2010}\\ 
\hline
\end{tabular}
\end{center}
\caption{Double $\Lambda$ binding energies $B_{\Lambda\Lambda}$ in unit of MeV calculated 
from theory \cite{gul2014} and measured from experiments \cite{naka2010,dany1963}. 
The SLy4 and HP$\Lambda$2 models are used for the $NN$ interaction and 
the $N\Lambda$ interaction, respectively. The results with \sllonep and \sllsamp are not presented 
because they are not available.}
\label{tab:lambinding}
\end{table}

In addition to the Skyrme-type force, we employ a finite-range force (FRF) model 
for the $\Lambda\Lambda$ interaction as given in Ref.~\cite{hiyama2002}. 
In this model, the potential is assumed in the form
\begin{equation}
v^{\mbox{\tiny FRF}}_{\Lambda\Lambda}(r) = \sum^3_{i=1} 
( v_i + v^\sigma_i \bm{\sigma}_\Lambda \cdot \bm{\sigma}_\Lambda ) e^{- \mu_i r^2}.
\end{equation}
The model parameters in the potential are fixed in order to reproduce $B_{\Lambda\Lambda}$
of $^6_{\Lambda\Lambda}{\rm He}$, and their values are listed in Table IV of Ref.~\cite{hiyama2002}.
We use the same values for our calculations. 

Because it is the first attempt to consider the non-Skyrme-type interactions 
in the framework of the Skyrme-type models, 
it is necessary to address the motivation for such an attempt.
The primary motivation is the fact that the predictions from the Skyrme-force models 
are not fully consistent with the experimental binding energies 
as shown in Table \ref{tab:lambinding}. 
For example, among the three S$\Lambda\Lambda$ models, the \sllsam model predicts best 
the binding energy of $^6_{\Lambda\Lambda}{\rm He}$, 
but its predicted values for the more massive nuclei deviate from the experimental values 
more severely than those predicted from the other two models. In contrast, the \sllone model 
behaves in the opposite way. For this reason, we decide to consider the finite-range force 
model whose predictions are in reasonable agreement with the experimental values 
(Table V in Ref.~\cite{hiyama2002}), although the applications of the finite-range force model 
were somehow limited. 
Furthermore, as mentioned in the introduction, the purpose of this work is to explore the effect of 
the $\Lambda\Lambda$ interactions more broadly and precisely, in order to solve the hyperon 
puzzle. Thus, it is worth considering as many models as possible that predict the experimental 
data consistently. 
Note that the hyperon-hyperon interactions in the RMF models 
are treated very naively, while several Skyrme-force models or few-body models, whose model parameters 
are adjusted explicitly to the experimental hypernuclear data, are available. Therefore, we 
believe that using these Skyrme-type force models or a finite-range force model is better to model 
the $\Lambda\Lambda$ interactions in combination with 
the Skyrme-type models for the $NN$ and $N\Lambda$ interactions.

In general, when the $\Lambda\Lambda$ interactions are introduced, 
the $N\Lambda$ interactions are usually fitted first with the background $NN$ interaction models, 
and then the $\Lambda\Lambda$ interactions are added on top of the $NN$ and $N\Lambda$ 
interaction models.
The model parameters of the $NN$, $N\Lambda$ and $\Lambda\Lambda$ interactions determined in this way 
form a single complete set of the interaction model in consideration.
In this respect, the way that we determine the model parameters in this work is not fully 
self-consistent because the ${\rm S}\Lambda\Lambda$ interaction 
models in Ref. \cite{lans1998} were determined with the SkM* 
for the $NN$ interaction and the YBZ5 for the $N\Lambda$ interaction
for the fitting of the $\Lambda\Lambda$ potential parameters. 
However, this approach may be still valid because Table \ref{tab:lambinding} shows that the predicted 
binding energies are in relatively good agreement with the experimental data in spite of some 
partial mismatches in the interaction models. 
Furthermore, because it is generally known that the EoS of NS matter is not quite 
sensitive to the fine tuning to individual nuclei, 
the resulting mass-radius relation is not affected much by the details of the fitting procedure.
For these reasons, we do not think that the inconsistency in the combined interactions that we used for 
this work will change the major conclusions of the final results.

Once all the interactions are fixed, it is straightforward to calculate their matrix elements 
in the uniform infinite matter. The Hamiltonian density for nucleons is obtained as 
(for the details of derivation, see, e.g. \cite{iwata2012})
\begin{eqnarray}
{\cal H}_N &=& \sum_{i=n,\, p}\frac{\hbar^2}{2 m_N}\tau_i + \rho_N (\tau_n + \tau_p)
\left[\frac{t_1}{4}\left(1+\frac{x_1}{2}\right) + \frac{t_2}{4}\left(1+\frac{x_2}{2}\right)\right]
\nonumber \\ &&
+ \sum_{i=n,\, p} \tau_i \rho_i 
\left[-\frac{t_1}{4}\left(\frac{1}{2}+x_1\right) + \frac{t_2}{4}\left(\frac{1}{2}+x_2 \right)\right]
\nonumber \\ &&
+ \frac{t_0}{2} \left[\left(1+\frac{x_0}{2}\right) \rho^2_N
- \left(\frac{1}{2}+x_0\right)(\rho^2_n + \rho^2_p)\right]
\nonumber \\ &&
+ \frac{t_3}{12} \left[\left(1+\frac{x_3}{2}\right) \rho^2_N
- \left(\frac{1}{2}+x_3\right)(\rho^2_n + \rho^2_p)\right] \rho^\epsilon_N,
\label{eq:hn}
\end{eqnarray}
and the terms for $\Lambda$ hyperon read
\begin{eqnarray}
{\cal H}_\Lambda &=&  \frac{\hbar^2}{2m_{\Lambda}}\tau_{\Lambda}
+ u_0 \Bigl(1+\frac{1}{2}y_0 \Bigr)\rho_N\rho_{\Lambda}
+ \frac{1}{4}(u_1 + u_2)(\tau_{\Lambda}\rho_N + \tau_N\rho_{\Lambda})
\nonumber \\ &&
+ \frac{3}{8}u_3^{\prime}\Bigl(1+\frac{1}{2}y_3\Bigr)\rho_N^{\gamma+1}
\rho_{\Lambda} + {\cal H}_{\Lambda\Lambda},
\label{eq:hl}
\end{eqnarray}
where $\tau_N = \tau_n + \tau_p$, and 
${\cal H}_{\Lambda\Lambda}$ stands for the term for the $\Lambda\Lambda$ interactions. 
The Skyrme-type force results in 
\begin{eqnarray}
{\cal H}_{\Lambda\Lambda} = \frac{\lambda_0}{4}\rho_{\Lambda}^2 
+ \frac{1}{8}(\lambda_1 + 3\lambda_2)\rho_{\Lambda}\tau_{\Lambda}
+ \frac{\lambda_3}{4}\rho_{\Lambda}^2\rho_N^{\alpha},
\label{eq:endenll}
\end{eqnarray}
and the finite-range force gives 
\begin{eqnarray}
{\cal H}^{\mbox{\tiny FRF}}_{\Lambda\Lambda} =
\frac{1}{2} \sum^3_{i=1} v_i \left(\frac{\pi}{\mu_i}\right)^\frac{3}{2} \rho^2_\Lambda
-\frac{1}{6 \pi^4} \sum^3_{i=1} (v_i + 3 v^\sigma_i)(\pi \mu_i)^\frac{3}{2} F(x_i),
\end{eqnarray}
where $x_i = k_F/\sqrt{\mu_i}$ and $F(x) = e^{-x^2} (x^2-2)-(3x^2-2)+\sqrt{\pi} x^3 {\rm erf}(x)$. 
Here, $k_F$ is the Fermi momentum and the error function, 
${\rm erf}(x) = \frac{2}{\sqrt{\pi}} \int^x_0 e^{-u^2} du$.


\section{Results}

The total energy density $({\cal E}= {\cal E}_N + {\cal E}_\Lambda + {\cal E}_e + {\cal E}_\mu)$ 
of the bulk matter can be obtained by minimizing the total Hamiltonian density 
(${\cal H} ={\cal H}_N + {\cal H}_\Lambda + {\cal H}_e + {\cal H}_\mu$) 
with the constraint of the charge neutrality condition, where ${\cal E}_N$ and
${\cal E}_\Lambda$ are evaluated from Eqs.~(\ref{eq:hn}, \ref{eq:hl}) 
and ${\cal E}_e$ and ${\cal E}_\mu$ can be found in Ref.~\cite{lim2014}.
Pressure is then calculated from the relation
\begin{equation}
P = \rho^2 \frac{\partial}{\partial\rho} \left(\frac{\cal E}{\rho}\right).
\end{equation}
NS matter satisfies baryon number conservation, charge neutrality, and $\beta$-equilibrium,
among which the second and third conditions read, respectively, 
\begin{equation}
\rho_p = \rho_e + \rho_\mu,
\end{equation}
and
\begin{equation}
\mun = \mup + \mue \,,
\quad \mue = \mumu\,,
\quad \mun + m_{n} = \mul + m_{\Lambda},
\label{eq:chemeq}
\end{equation}
where the chemical potential is obtained from 
\begin{equation}
\mu_i = \frac{\partial {\cal E}}{\partial \rho_i}.
\end{equation}

By solving the above equations self-consistently, we can determine the EoS. Then,  
by plugging the EoS into the TOV equation, we obtain the maximum mass and the 
mass-radius relation of NS. Because our main focus is to solve the
hyperon puzzle which is related to the maximum mass, we first present 
the results of the maximum mass obtained from our selected models of 
the $NN$, $N\Lambda$, and $\Lambda \Lambda$ interactions. We find that 
some of our selected models can predict the observed maximum mass of $\sim 2 M_\odot$, which 
provides a clue to solve the hyperon puzzle. Because it is necessary to investigate 
these models in more detail, we also present the EoS and the $\Lambda$ fraction 
obtained from these models. Finally, the mass-radius relation predicted from these 
models is compared with the observation.

\subsection{Maximum mass of NS}

\begin{table}[tbp]
\begin{center}
\begin{tabular}{c|c|c|c|c|c|c|c|c}
\hline 
 {~~$NN$ model~~} &  \multicolumn{2}{c|}{SLy4} & \multicolumn{3}{c|}{SkI4}  &  \multicolumn{3}{c}{SGI}     \\ [-1.5ex]
  {~(without $\Lambda$)$^\dagger$ } &  \multicolumn{2}{c|}{(2.07)} &   \multicolumn{3}{c|}{(2.19)}  &  \multicolumn{3}{c}{(2.25)}   \\
\hline 
 {$N\Lambda$ model}  & ~HP$\Lambda 2^\star$~ & ~~O$\Lambda 2$~~  
   & ~HP$\Lambda 2$~ 
   & ~~O$\Lambda 2$~~  
   & ~YBZ6$^\star$~ 
   & ~HP$\Lambda 2$~ 
   & ~~O$\Lambda 2$~~ 
   & ~YBZ6$^\star$~ \\  [-1.5ex]

  {~~(no $\Lambda\Lambda$)$^{\dagger\dagger}$}
   & (1.51)
   & (1.08)
   & (1.52)
   & (1.19)
   & (1.80)
   & (1.52)
   & (1.22)
   & (1.79)  \\ 
\hline

\sllone
   &  {{\it 1.40}} 
   &  {1.00}
   &  {1.41}
   & {1.12}
   & {\it 1.70}
   &  {1.42}
   &  {1.16 }
   & {\it 1.69}  \\
 
\slltwo
   &  {{\it 1.58}} 
   &  {1.28}
   &  {1.57}
   &  {1.30}
   & {\it 1.79}
   &  {1.57}
   &  {1.28}
   & {\it 1.77} \\ 

 
\sllsam
   &  {{\it 1.85}} 
   &  {1.57}
   &  {1.87}
   &  {1.62}
   & {\it 2.03}
   &  {1.88}
   &  {1.65}
   & {\it 2.04} \\ 

 
\sllonep
   & \multicolumn{1}{c| }{{\it 1.51}} 
   & \multicolumn{1}{c| }{1.08}
   & \multicolumn{1}{c| }{1.51 }
   &\multicolumn{1}{c| }{1.18}
   & {\it 1.79}
   & \multicolumn{1}{c| }{1.51}
   & \multicolumn{1}{c| }{1.21}
   & {\it 1.78} \\ 

 
\sllsamp
   & \multicolumn{1}{c| }{{\it 1.76}} 
   & \multicolumn{1}{c| }{1.43}
   & \multicolumn{1}{c| }{1.76}
   &\multicolumn{1}{c| }{1.47}
   & {\it 1.97}
   & \multicolumn{1}{c| }{1.77}
   & \multicolumn{1}{c| }{1.49}
   & {\it 1.96} \\ 

 
FRF
   & \multicolumn{1}{c| }{{\it 1.61}} 
   & \multicolumn{1}{c| }{1.22}
   & \multicolumn{1}{c| }{1.60}
   &\multicolumn{1}{c| }{1.25}
   & {\it 1.86}
   & \multicolumn{1}{c| }{1.59}
   & \multicolumn{1}{c| }{1.26}
   & {\it 1.84} \\ 
\hline
\end{tabular}
\\[1.0ex]
$^\dagger$  Maximum NS mass without both $N\Lambda$ and $\Lambda\Lambda$ interactions.\\[-1.5ex]
$^{\dagger\dagger}$  Maximum NS mass with $N\Lambda$ but without $\Lambda\Lambda$ interactions.
\end{center}

\caption{Maximum mass of NS in units of solar mass with the $NN$, $N\Lambda$, 
and $\Lambda\Lambda$ interactions.  $^\star$ Numbers in italic correspond to three selected models,
SLy4-HP$\Lambda$2, SkI4-YBZ6, and SGI-YBZ6, which give relatively large maximum NS mass.}
\label{tab:nlmass}
\end{table}

We present the maximum mass of NS for our selected models in Table \ref{tab:nlmass}. 
In order to see the effect of the $\Lambda$ hyperon, 
we considered three models for the $N\Lambda$ interaction (HP$\Lambda2$, O$\Lambda2$, and YBZ6) and
six models for the 
$\Lambda\Lambda$ interaction.
%
%
As expected, including only the $N\Lambda$ interactions on top of the $NN$ interactions 
reduces the maximum mass of NS significantly.
It is also shown that the extent of the decrease in the maximum mass 
strongly depends on the choice of the $N\Lambda$ interaction. For the same $NN$ interaction model, 
the maximum mass is the smallest with the O$\Lambda2$ model, while it is the largest with 
the YBZ6 model. This implies that the relative stiffness of the $N\Lambda$ interactions 
increases in the order of O$\Lambda2$, HP$\Lambda2$, and YBZ6. 
The decrease of the maximum mass due to the $N\Lambda$ interaction 
ranges from $0.4 M_\odot$ to $1.0 M_\odot$. Note that this range that we obtained 
is much wider than that predicted from the RMF models, which is $0.2 M_\odot - 0.5 M_\odot$ 
(Ref.~\cite{weiss2012} and references therein).
Our result (predicting a wider range of the decrease in the maximum mass)
is very interesting because all the three $N\Lambda$ models that we considered give very similar
values for the optical potential depth at the saturation density (Table \ref{tab:nlam}).
Without including the $\Lambda\Lambda$ interaction, the O$\Lambda2$ model predicts too small values for 
the maximum mass regardless of the choice of the $NN$ interaction models to be consistent with the 
lower limit of the canonical range of NS mass. Thus, the O$\Lambda2$ model may be ruled out in the 
case that the effect of the $\Lambda$ hyperon is only the $N\Lambda$ interaction. 
Note that the result for YBZ6 in combination with SLy4 is not presented because this case is not 
physically plausible. In this case, $\mu_\Lambda$ is large and it also increases 
faster than $\mu_n$, making the chemical equilibrium between $\Lambda$ and neutron 
unreachable, i.e., there is no solution to the third equation of Eq.~(\ref{eq:chemeq}). 
This may imply that the creation of hyperons in NS is possible only when the interplay 
between the $NN$ and $N\Lambda$ interactions is optimized.

Unlike the $N\Lambda$ interactions that always reduce the maximum mass when they are 
included on top of the $NN$ interactions, the $\Lambda\Lambda$ interactions can 
both increase and decrease the maximum mass when they are included on top of 
the $NN$ and $N\Lambda$ interactions. However, the effect of the $\Lambda\Lambda$ interaction 
on the maximum mass varies with the selection of the model. The S$\Lambda\Lambda1$ and 
S$\Lambda\Lambda 1'$ models do not increase the maximum mass regardless of 
the $NN$ and $N\Lambda$ interaction models that we considered. 
In contrast, the \sllsam, \sllsamp, and FRF models always increase the maximum mass
with such a trend that the amount of mass increase from the $N\Lambda$ value 
increases in the order of FRF, \sllsamp, and \sllsam for the given $NN$ and $N\Lambda$ 
interactions. Meanwhile, the \slltwo model either increases or decreases the maximum 
mass depending on the choice of the $NN$ and $N\Lambda$ interaction models. 
Note that the changes in the maximum mass from their $N\Lambda$ values 
due to the inclusion of the various $\Lambda\Lambda$ interactions in our consideration 
range from $-0.1 M_\odot$ to $0.4 M_\odot$. This shows that the effect of the  $\Lambda\Lambda$ 
interactions on the maximum mass obtained from our calculations is larger than that obtained 
from the RMF models in which the $\Lambda\Lambda$ interactions change the maximum mass 
from the $N\Lambda$ values only by $\pm 0.1 M_\odot$ \cite{bed2005, mi2010}.

The most important result that comes from Table \ref{tab:nlmass} is that some of 
our models that combine the suitable $NN$, $N\Lambda$, and $\Lambda\Lambda$ interactions 
predict large values for the maximum mass of NS 
even though the hyperons are included. For example, the \sllsam and \sllsamp models predict 
the recently observed maximum mass of NS, $\sim 2 M_\odot$, when they are combined 
with the SkI4-YBZ6 and SGI-YBZ6 models. Because these models can provide a clue to solve 
the hyperon puzzle, it is worth investigating these models in more detail. 
For this reason, we select some models from Table \ref{tab:nlmass} which predict the large 
maximum mass only and present the EoS and the $\Lambda$ fraction obtained from these models 
in the next two sections. 

\subsection{Equation of state}

\begin{figure}
\begin{center}
\includegraphics[scale=0.8]{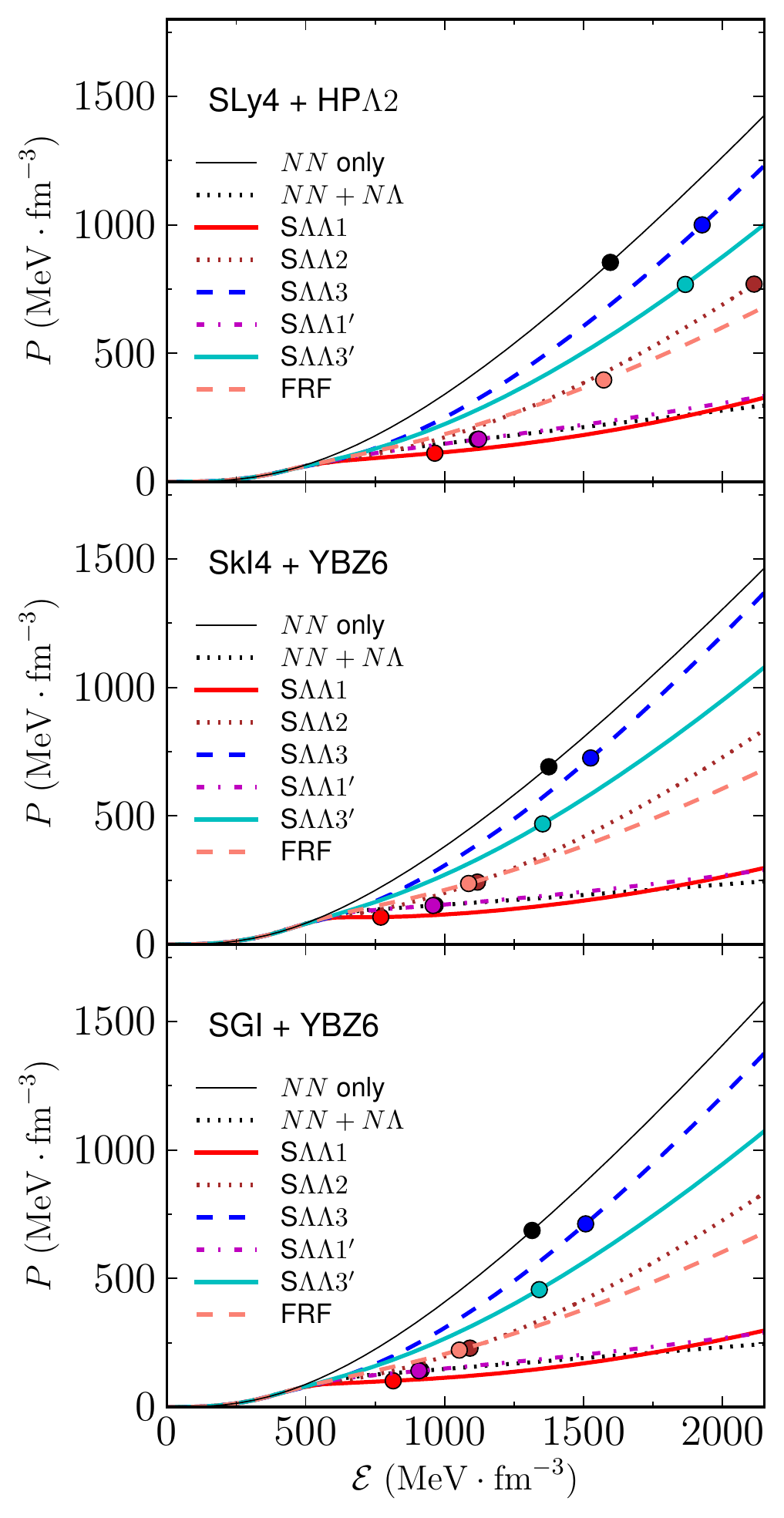} 
\end{center}
\caption{(Color online) EoS for the selected models that predict the large maximum mass 
of NS. Different lines indicate the various $\Lambda\Lambda$ interactions that we 
consider in this paper. For comparison, the EoS without any contribution 
from the $\Lambda$ hyperon ($NN$ only) is also plotted. 
Filled circles indicate the locations on the EoS curves 
corresponding to the center of NS with the maximum mass.}
\label{fig:eos}
\end{figure}

Figure \ref{fig:eos} shows the EoS of NS matter for the selected models that 
give large values for the maximum mass of NS. By considering the slopes in 
the plots, we can sort out the models into three kinds; 
\sllone and \sllonep as soft, \slltwo and FRF as mild, and \sllsam and \sllsamp as hard.

In general, models with stiffer EoS predict larger maximum mass and our results in Table~\ref{tab:nlmass}
show that 
most of the models follow this general trend. However, an exception to this general 
trend seems to be seen in \slltwo in comparison with \sllonep and FRF, in particular 
when we consider the slopes of the EoS curves at high densities (correspond to large 
${\cal E}$ in the plots). 
The EoS curve of \slltwo (dotted line) is much stiffer than that of \sllonep (dot-dashed line), 
especially at high densities, but the maximum mass predicted 
from \slltwo is just slightly larger or 
even smaller than that predicted from \sllonep. 
Similarly, FRF (orange dashed line) is softer than \slltwo also at high densities, 
but the maximum mass predicted from FRF 
is larger than that predicted from \slltwo for all three cases in consideration.

In order to understand this exceptional behavior between the EoS and the maximum mass 
seen in some models, we mark the locations 
corresponding to the center of NS 
with the maximum mass by filled circles on the EoS curves in Figure \ref{fig:eos},
and summarize the densities at the center of maximum-mass stars in Table \ref{tab:center}.
\begin{table}
\begin{center}
\begin{tabular}{cccc}\hline
  &SLy4+HL$\Lambda$2 & SkI4 + YBZ6 & SGI + YBZ6 \\ \hline
Without $\Lambda\Lambda$ & 1.00 & 0.85 &  0.81 \\ 
S$\Lambda\Lambda1$ & 0.89 & 0.71 & 0.74 \\ 
S$\Lambda\Lambda2$ & 1.65 & 0.96 & 0.94 \\ 
S$\Lambda\Lambda3$ & 1.44 & 1.17 & 1.15 \\ 
S$\Lambda\Lambda1'$ & 1.00 & 0.85 & 0.81 \\ 
S$\Lambda\Lambda3'$ & 1.44 & 1.09 & 1.08 \\ 
FRF                               & 1.31 & 0.93 & 0.91  \\ \hline
\end{tabular}
\end{center}
\caption{
Density in fm$^{-3}$ at the center of stars corresponding to 
the locations of filled circles in Fig.~\ref{fig:eos}.}
\label{tab:center}
\end{table} 
First of all, the locations of filled circles vary significantly from
model to model, which 
implies that NS having the maximum mass in each model has different energy density 
and pressure at the center. Note that the origin of the plot corresponds to the 
surface of NS and the EoS curve traces the interior of NS (as a function of 
radius) from the surface to the center 
corresponding to the filled circle. 
Roughly speaking, the maximum mass is obtained by integrating the energy density 
over the entire NS, more accurately over the radius of NS in 1D spherical 
coordinates. Therefore, the part of the EoS curve only between the origin and the 
filled circle is relevant to the calculation of the maximum mass. Having this in mind, 
we re-examine the three EoS curves of \slltwo, \sllonep, and FRF, and find that within
the relevant part of the EoS curve, the stiffness-maximum mass trend is not violated 
severely. FRF is always stiffer than \slltwo at low densities, which is consistent with 
the results of the maximum mass, i.e., the maximum mass of FRF is always larger than 
that of \slltwo in all three cases in consideration. In case of SLy4+HP$\Lambda2$, 
the relevant EoS curve of \slltwo is extended to larger energy density than that of 
FRF, implying that the calculation of the maximum mass should include 
the contribution from the partial EoS curve of \slltwo 
between ${\cal E} \approx 1600 \rm{MeV} \cdot \rm{fm}^{-3}$ and 
${\cal E} \approx 2200 \rm{MeV} \cdot \rm{fm}^{-3}$ in comparison with the relevant 
EoS curve of FRF. 
However, the contribution from this partial EoS curve of \slltwo 
to the maximum mass is only $0.01 M_\odot$, so one can deduce that this part
is so close to the center of the NS that it occupies only a small fraction of the total volume, 
and thus the total mass.
In other words, in the \slltwo model, the energy density changes quite a lot 
near the center in comparison with the FRF model. 
Similarly, the EoS curve of \sllonep
that is relevant to the calculation of the maximum mass is as steep as that of \slltwo, 
in particular at low energy densities (between $\sim 500 \rm{MeV} \cdot \rm{fm}^{-3}$ 
and $\sim 750 \rm{MeV} \cdot \rm{fm}^{-3}$) for the cases of SkI4+YBZ6 and SGI+YBZ6. Thus, 
the stiffness-maximum mass trend also holds between \sllonep and \slltwo. 

A brief summary of this section is as follows. 
For all of our selected models that predict 
the large maximum mass, a general trend between the stiffness of the EoS 
and the maximum mass of NS (i.e., the stiffer the EoS, the larger the maximum 
mass) always holds. Because all of our selected models include 
all the interactions involving the $\Lambda$ hyperons 
($NN$, $N\Lambda$, and $\Lambda\Lambda$), we can say 
that the general trend holds even when the full effect of the $\Lambda$ hyperons 
is included. The $\Lambda$ hyperons affect the EoS in such a way that the more 
the $\Lambda$ hyperons exist, the softer the EoS becomes. So, in the following 
section, we present the fraction of the $\Lambda$ hyperons for the same set of 
our selected models.

\subsection{$\Lambda$ hyperon fraction}

\begin{table}
\begin{center}
\begin{tabular}{c|cccccccc}
\hline
 & \multicolumn{2}{c|}{SLy4} &  \multicolumn{3}{c|}{SkI4} & \multicolumn{3}{c}{SGI} \\
\cline{2-9}
&\multicolumn{1}{c|}{~~HP$\Lambda 2$~~} & \multicolumn{1}{c|}{~~O$\Lambda 2$~~}
& \multicolumn{1}{c|}{~~HP$\Lambda 2$~~} & \multicolumn{1}{c|}{~~O$\Lambda 2$~~}
& \multicolumn{1}{c|}{~~YBZ6~~} 
& \multicolumn{1}{c|}{~~HP$\Lambda 2$~~} & \multicolumn{1}{c|}{~~O$\Lambda 2$~~}
& \multicolumn{1}{c}{~~YBZ6~~} \\
\hline
$\rho_{\rm crit}$ 
&\multicolumn{1}{c|}{0.453} 
& \multicolumn{1}{c|}{0.380}
& \multicolumn{1}{c|}{0.374}
& \multicolumn{1}{c|}{0.340}
& \multicolumn{1}{c|}{0.455} 
& \multicolumn{1}{c|}{0.352}
& \multicolumn{1}{c|}{0.325}
& \multicolumn{1}{c}{0.412} \\
\hline
$M/M_\odot$
&\multicolumn{1}{c|}{1.17} 
& \multicolumn{1}{c|}{0.90}
& \multicolumn{1}{c|}{1.19}
& \multicolumn{1}{c|}{1.04}
& \multicolumn{1}{c|}{1.51} 
& \multicolumn{1}{c|}{1.21}
& \multicolumn{1}{c|}{1.08}
& \multicolumn{1}{c}{1.47} \\
\hline

\end{tabular}
\end{center}
\caption{The critical density $\rho_{\rm crit}$ in units of fm$^{-3}$ and the corresponding 
mass of NS that has the central density equal to the critical density.}
\label{tab:ncritical}
\end{table}

In Table \ref{tab:ncritical}, we list the critical density $\rho_{\rm crit}$ at which 
hyperons begin to appear in the interior of NS and the corresponding mass of NS in units of 
solar mass ($M/M_\odot$) whose central density is $\rho_{\rm crit}$. 
The critical density increases with the stiffness of the $N\Lambda$ interaction, in such a 
way that YBZ6 $>$ HP$\Lambda2$ $>$ O$\Lambda2$ for a given $NN$ interaction. 
This tendency can be understood from Eq. (\ref{eq:chemeq}); more repulsive $N\Lambda$ interactions 
result in larger chemical potentials of $\Lambda$ hyperons. 
In contrast, for a given $N\Lambda$ interaction, the critical density decreases with the stiffness 
of the $NN$ interaction (the $NN$ interaction models get stiffer in the way that SGI $>$ SkI4 
$>$ SLy4). This can also be understood from Eq. (\ref{eq:chemeq}); 
more repulsive $NN$ interactions result in large chemical potentials of nucleons ($\mu_n$) and 
make the chemical equilibrium between nucleons and $\Lambda$ hyperons reached at the lower density. 

\begin{figure}
\begin{center}
\includegraphics[scale=0.8]{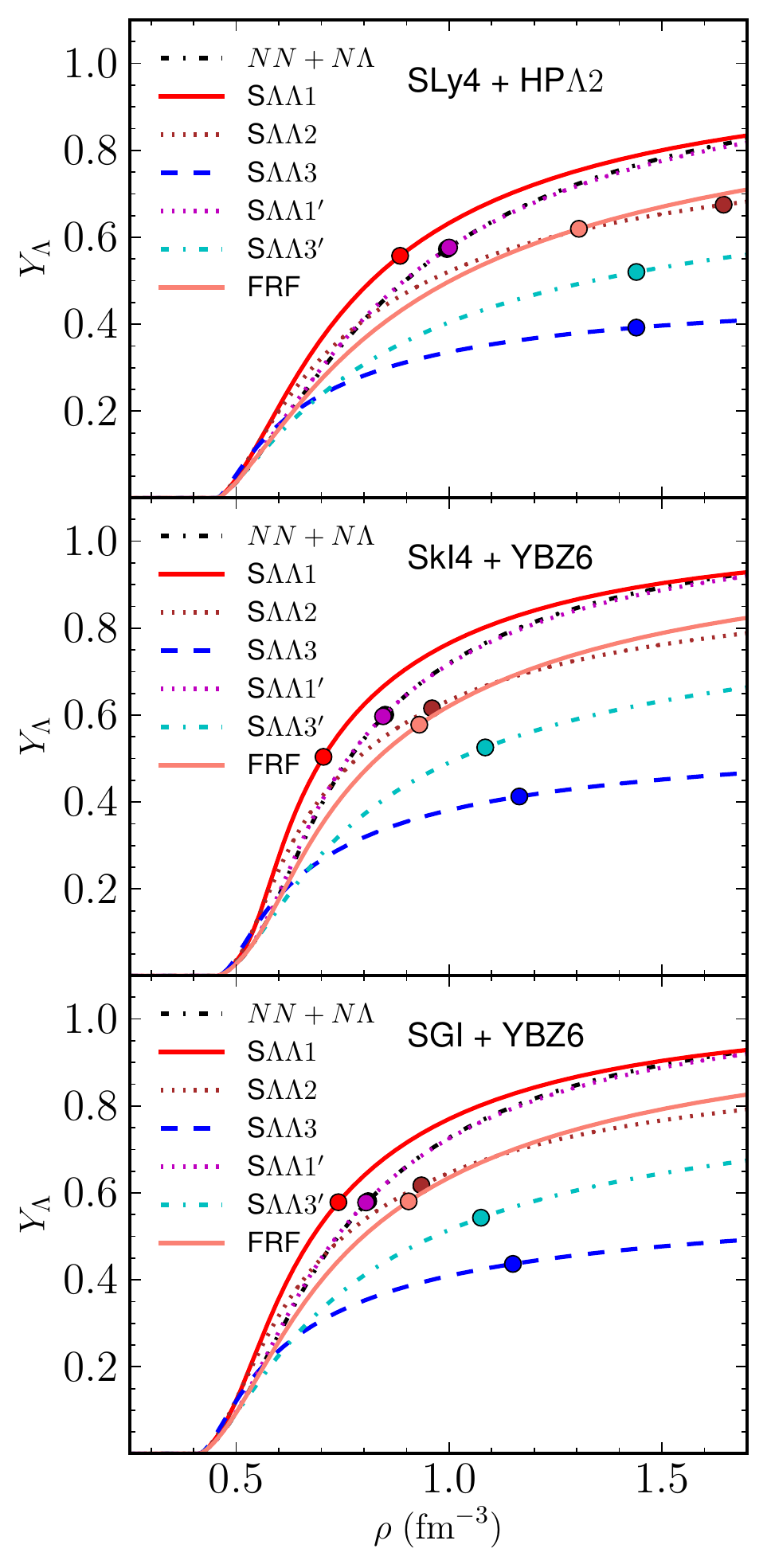} 
\end{center}
\caption{(Color online) Particle fraction of $\Lambda$ hyperons as a function of the total 
baryon number density. The $\Lambda$ fraction ($Y_{\Lambda}$) is defined as $\rho_\Lambda / \rho$, 
where $\rho$ is the total baryon number density, i.e., $\rho = \rho_n + \rho_p + \rho_\Lambda$.}
\label{fig:frac}
\end{figure}

Figure \ref{fig:frac} displays the fraction of $\Lambda$ hyperons as a function of the total baryon number density
 for the same selected models for which the EoS was presented in 
Figure \ref{fig:eos}. Regardless of the models, the $\Lambda$ fraction behaves very 
similarly near the critical density, but depending on the $\Lambda\Lambda$ 
interaction, it varies significantly as the total baryon number density increases. 
As in Figure \ref{fig:eos}, we mark the locations of the center of NS with the maximum 
mass by using filled circles in Figure \ref{fig:frac}. Note again that the part of the 
fraction curve after the filled circle (i.e., the fractions at the large total baryon densities)
is not relevant to the calculation of the maximum mass. 
Up to the locations of the filled circles, models with smaller $\Lambda$ fractions
predict the larger maximum mass of NS. In particular, 
the particle fraction curves of 
\slltwo, \sllonep, and FRF (for which we discussed the general trend between the EoS and 
the maximum mass in more detail in the previous section) 
show distinctively the trend between the fraction and the maximum 
mass. The fraction of FRF is always smaller than that of \slltwo in the relevant part of 
the fraction curve, which is consistent with the trend of the maximum mass such that 
the maximum mass of FRF is always larger than that of \slltwo. Between \slltwo and \sllonep, 
the fraction curves of these two models cross each other before they reach 
the locations of filled circles, which implies that the fraction of $\Lambda$ hyperons 
in \sllonep is very similar to that in \slltwo up to the locations of the filled circles 
in the sense of average (although the fraction curves of 
\sllonep are always above 
those of \slltwo at large total baryon densities).  

Combining the two trends, one between the EoS and the maximum mass shown in Figure 
\ref{fig:eos} and the other between the $\Lambda$ fraction and the maximum mass 
shown in Figure \ref{fig:frac}, we can confirm the effect of $\Lambda$ hyperons 
on the EoS; the more $\Lambda$ hyperons exist inside NS, the softer the EoS 
becomes (thus making the maximum mass smaller). 
However, the discussions so far, which are based on the maximum mass, the EoS, and 
the $\Lambda$ fractions, do not include any information on the size of NS which 
also constrains the model predictions together with the mass. In the next section, 
we present the mass-radius relation obtained from our selected models that predict 
the large maximum mass.

\subsection{Mass-radius relation}

\begin{figure}
\begin{center}
\includegraphics[scale=0.8]{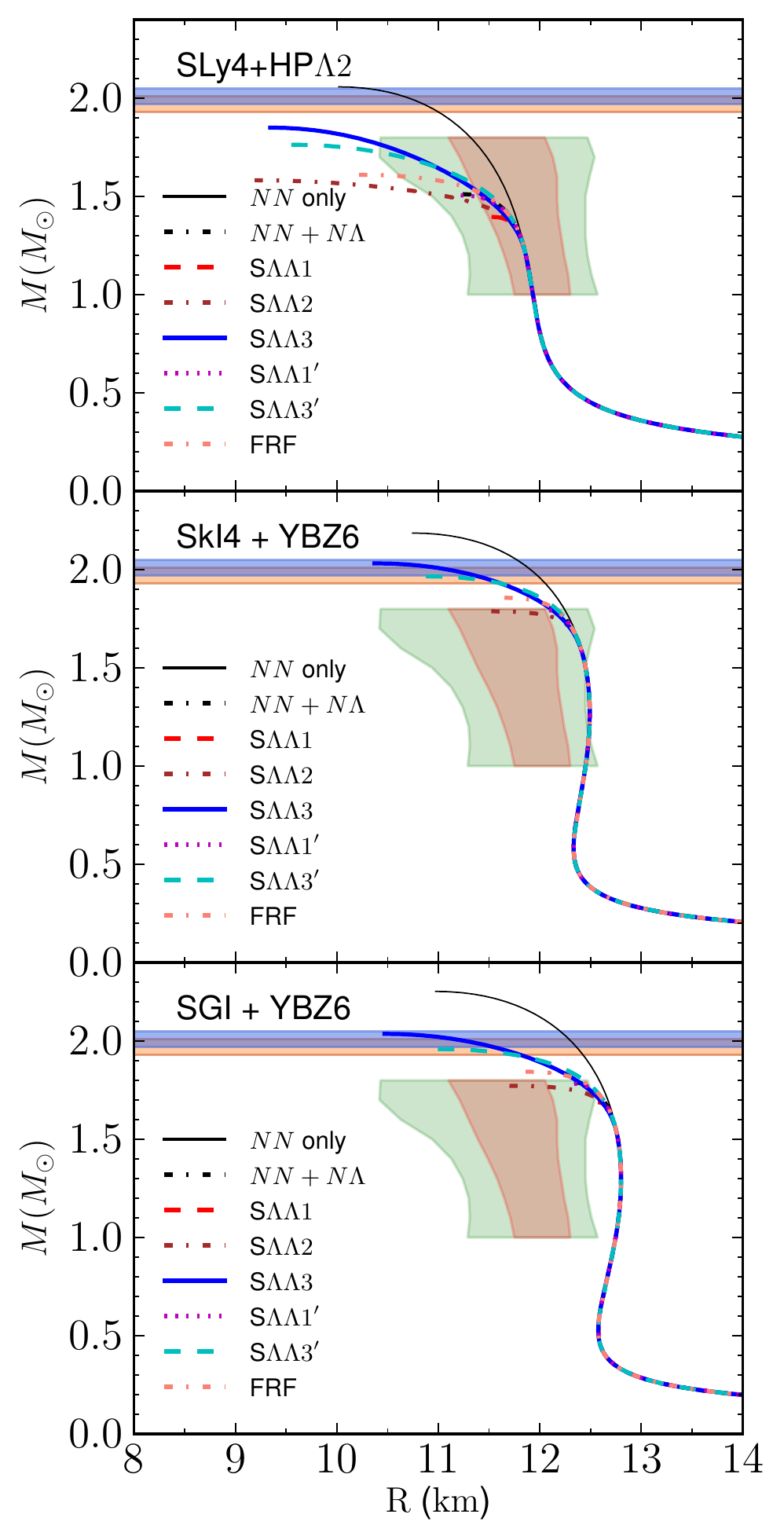} 
\end{center}
\caption{(Color online) Mass-radius of neutron stars for the selected 
models. For comparison, the mass-radius relation without any contribution 
from $\Lambda$ hyperons is also plotted as a black solid line in each 
panel. 
Two flat bands (with purple and orange color) denote the mass ranges
of PSR J1614-2230 ($1.97 \pm 0.04 M_\odot$) and 
PSR J0348+0432 ($2.01\pm0.04 M_\odot$), respectively. 
Green and brown regions near the center are the mass-radius
ranges obtained from Ref. \cite{steiner2010}.}
\label{fig:tov}
\end{figure}
Figure \ref{fig:tov} shows the mass-radius relation of neutron stars
for the selected models 
which were used to get the EoS and $\Lambda$ fractions in
Figures~\ref{fig:eos} 
and \ref{fig:frac}, respectively.
Flat bands around $2 M_\odot$ mark the range $(1.97\pm0.04) M_\odot$ and
$(2.01\pm0.04) M_\odot$, masses of PSR J1614-2230 and PSR J0348+0432, respectively.
Green and brown regions (color online) near the center of each panel
represent the mass-radius constraints obtained from statistical analysis of the
observed NS in Ref. \cite{steiner2010}. 
For simplicity, we call the former constraint PSR1614 and the latter SLB2010.
The mass-radius relation justifies our selection of the three $NN$ models, 
which are in reasonable agreement with both the PSR1614 and SLB2010 constraint. 
A slight deviation is seen in case of SGI, in which the mass-radius curve of SGI  
(the black solid line in the bottom panel of Figure~\ref{fig:tov}) 
is outside the SLB2010 zone by about 0.5~km in the radius. However, under the 
large uncertainties in the observed size of NS, we believe that it is worth 
considering the models with the 0.5~km deviation in the radius such as SGI.

As expected, at the same radius, the $\Lambda$ hyperons reduce the mass 
of NS in comparison with the case without them. 
This is shown in Figure \ref{fig:tov}. Furthermore, the reduction in the mass 
due to the $\Lambda$ hyperons (from the case without them) decreases with 
the stiffness of the EoS. In other words, when the hyperons are included, 
the models with stiff EoS (such as S$\Lambda\Lambda$3) have larger mass than 
those with soft EoS (such as \sllone) at the same radius. 
However, even when the hyperons are included, the models in our consideration
can still satisfy both the PSR1614 and SLB2010 constraint, which supports the 
conclusion that we drew for the hyperon puzzle before; the maximum mass 
of $\sim 2 M_\odot$ is achievable even with the $\Lambda$ hyperons included. 
The models that predict the maximum mass of $\sim 2 M_\odot$ are also in 
reasonable agreement with the mass-radius relation SLB2010, which puts an additional 
strong constraint on the models. Among our selected models, it is S$\Lambda\Lambda3$ 
and S$\Lambda\Lambda3'$ in combination of SkI4+YBZ that satisfy 
both the PSR1614 and SLB2010 best. The other models satisfy only 
one of these two constraints.

\section{SUMMARY AND DISCUSSION}

In this work, we try to solve the hyperon puzzle, which has been stated as that 
hyperons appearing in NS are not able to predict NS with large mass, 
especially $\sim 2 M_\odot$. In order to do this, we included the $\Lambda\Lambda$ interactions 
on top of the $N\Lambda$ and $NN$ interactions. In the previous studies such as the RMF models, 
the interactions between $\Lambda$ hyperons have been treated based upon the 
symmetry properties of quark models, thus the connection with the empirical data has 
not been so tight. Here, we employed $\Lambda\Lambda$ interactions 
which are adjusted to various data such as single-particle energy levels, binding energies, 
bond energies, and etc. of known double-$\Lambda$ hypernuclei. We find that the employed 
$\Lambda\Lambda$ interactions result in a wide variety of the maximum mass of NS, 
the EoS of NS matter, the $\Lambda$ fractions inside NS, and the mass-radius relation. 
However, some models satisfy both the recent constraint on the maximum mass ($\sim 2 M_\odot$)
and the mass-radius relation, which provides a clue to solve the hyperon puzzle, i.e., 
the existence of hyperons inside NS can be still compatible with the large-mass NS, 
even up to $\sim 2 M_\odot$.

\begin{figure}
\begin{center}
\includegraphics[scale=0.5]{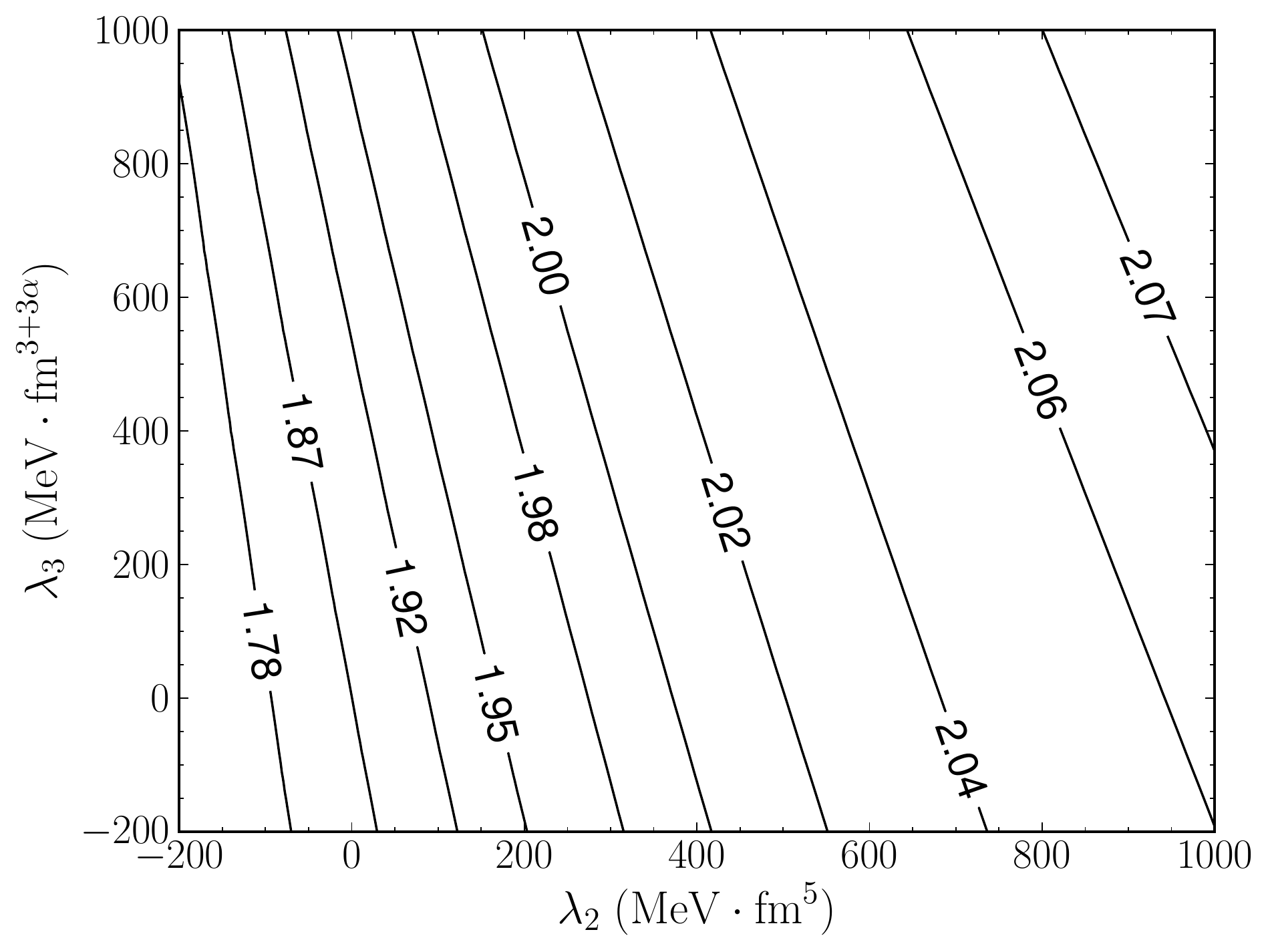}
\end{center}
\caption{Contour plot of the maximum mass of NS in units of solar mass
as functions of $\lambda_2$ and $\lambda_3$ which are used 
in the Skyrme-type $\Lambda\Lambda$ interaction models. Here, $\alpha$ is fixed to 1/3 
(see Table \ref{tab:lamlam}). The $NN$ and $N\Lambda$ interactions are SkI4 and HP$\Lambda2$, respectively. For $\Lambda\Lambda$ interation, S$\Lambda\Lambda3$
is used.}
\label{fig:contour}
\end{figure}

An important point of this work is that the models we adopted for the $NN$, $N\Lambda$, and 
$\Lambda\Lambda$ interactions are consistent with the currently 
available nuclear and hypernuclear data, thus additional {\it ad hoc} assumptions or modification 
of the model parameters are not required in order to make the hyperon stars feasible.
Note that uncertainties in the model predictions are primarily caused by fitting 
the model parameters to nuclear data with some uncertainties or just by lacking the sufficient data of 
double-$\Lambda$ hypernuclei. As an alternative approach, 
if we were to take into account the constraints from astronomical observations 
such as mass, mass-radius relation, and surface temperature of NS, we could possibly constrain 
the model parameters involved in the $\Lambda$ hyperons and obtain 
a better understanding of the state of matter at supra-saturation densities.
As an example of this approach, in Figure \ref{fig:contour},  
we plot the maximum mass of NS for certain ranges of 
$\lambda_2$ and $\lambda_3$ values which are used for the Skyrme-type $\Lambda\Lambda$ interaction
models as in Table \ref{tab:lamlam}.  
In this example, the combined model of SkI4 with HP$\Lambda$2 and S$\Lambda\Lambda3$
is used to see the behavior the maximum mass of NS when
the variation of $\lambda_2$ and $\lambda_3$ are allowed. 
As seen in table \ref{tab:nlmass}, the maximum mass of the model is
only $1.87 M_\odot$ without the $\lambda_2$ momentum interaction and $\lambda_3$ density dependent
manybody interaction. The maximum mass
of NS, however, can reach up to $2.0 M_\odot$ if we consider 
repulsive $\lambda_2$ and $\lambda_3$. This also suggests the clue for
the hyperon puzzle.
When more abundant data for the double-$\Lambda$ hypernuclei are available, 
we may be able to determine the values
of $\lambda_2$ and $\lambda_3$ from the data and to compare them with the observations of the 
NS. 

Recently, authors of Ref. \cite{bs2015} calculated the energy of $\Lambda$ hyperons in nuclear matter
using a phenomenolocal formula, and obtained the distribution of parameters satisfying $M > 2 M_\odot$ 
with Monte Carlo simulation.
Aside from the experimental data, the constraint obtained in the work can be used as inputs for 
determining the parameters entering the interaction of $\Lambda$ hyperons.
The work of refitting is beyond the scope of this work, but it deserves a detailed analysis 
for better understanding of the interactions of $\Lambda$ hyperons.

As a final remark, we discuss some limitations of our present work. 
As mentioned before, the parameters of the $N\Lambda$ and $\Lambda\Lambda$ interactions 
are generally determined on top of a given
$NN$ interaction model, so the parameters of the whole interactions determined in this way 
form a unique set of an interaction model.
On the other hand, our investigations are based on a hybrid scheme, in which we combine the 
$NN$, $N\Lambda$, and $\Lambda\Lambda$ interactions piece by piece 
among different models.
For this reason, errors can be generated by the lack of self-consistency. 
%
\begin{table}
\begin{center}
\begin{tabular}{ccccc||cccc}
\hline
$NN$ & SLy10  & SV  & SLy4 & SGI  & $NN$ & SkM$^*$  & SkI4  & SGI  \\ [-1.5ex]
 &  (1.99) &  (2.43) & (2.07) &  (2.25) &  &  (1.62) &  (2.19) &  (2.25) \\
$N\Lambda$ & LY-I  & YBZ6  & HP$\Lambda2$  & YBZ6  & $N\Lambda$ & YBZ4  & 
YBZ6  & YBZ6  \\  [-1.5ex]
 &  (1.32) &  (1.66) &  (1.51) &  (1.79) &  &  (1.17) & 
 (1.52) &  (1.79) \\
Ref. & [27] & [27] & this work & this work & Ref. & [20] & this work & this work \\
\hline
$\Delta M1$ & $-0.11$ & $-0.10$ & $-0.11$ & $-0.10$ & $\Delta M1'$ & 0.01 & $-0.01$ & $-0.01$ \\
$\Delta M2$ & 0.11 & $-0.02$ & 0.07 & $-0.02$ & $\Delta M2'$ & - & - & - \\
$\Delta M3$ & 0.34 & 0.32 & 0.34 & 0.25& $\Delta M3'$ & 0.21 & 0.17 & 0.17 \\
\hline
\end{tabular}
\end{center}
\caption{NS maximum mass difference between with and without $\Lambda\Lambda$ interactions,
i.e. $\Delta Mn \equiv M(NN+N\Lambda+S\Lambda\Lambda n) - M(NN+N\Lambda)$.
All the values are in units of $M_\odot$, and the numbers in the parentheses denote the values 
with only $NN$ interactions in the row of $NN$, and with $NN+N\Lambda$ interactions in the row of $N\Lambda$.}
\label{tab:diff}
\end{table}

In order to see the effect of inconsistent combination of interaction models, we evaluate the
difference of NS maximum mass with and without $\Lambda\Lambda$ forces.
Table~\ref{tab:diff} summarizes the result where $\Delta M n$ ($\Delta M n'$) is defined
as the difference of NS maximum mass between with and without S$\Lambda\Lambda n$
(S$\Lambda\Lambda n'$) interactions. 
Parentheses denote the NS maximum mass for given $NN$ and $NN+N\Lambda$ interactions.
All the values in the table are in the unit of solar mass $M_\odot$.
$NN$ force models give significant variation for the NS maximum mass,
and the selected $N\Lambda$ forces show the reduction of the maximum mass in the 
range of $(0.46 \sim 0.77) M_\odot$.
Inspite of this strong dependence on the $NN$ and $N\Lambda$ interactions, $\Delta Mn$ and $\Delta Mn'$ 
are lesser sensitive to the $\Lambda\Lambda$ force models. 
We note that the S$\Lambda\Lambda n$ models are fitted to SkM$^*$ and YBZ5 models for
$NN$ and $N\Lambda$ interactions, respectively, so all the $\Delta M n$ values are results of
inconsistent model combinations.
Nevertheless, the contribution of $\Lambda\Lambda$ interaction is not
much affected by the inconsistency.
Right part of the table presents the result of $\Delta M n'$.
In Ref.~\cite{minato2011}, SkM$^*$ and YBZ4 models are used to fit the $\Lambda\Lambda$ interactions,
so $\Delta M n'$ values in the column of SkM$^*$ are the results of consistent use of interaction models.
Similar to $\Delta M n$, NS maximum mass shows dependence on the $\Lambda\Lambda$ interaction
at the order of $10^{-2} M_\odot$.
The reason for this weak dependence on the $\Lambda\Lambda$ interaction can be found from the
energy density in Eq.~(\ref{eq:endenll}), and the fraction of $\Lambda$ hyperons in Fig.~\ref{fig:frac}.
First, since $\lambda_2 = 0= \lambda_3$ in the $\Lambda\Lambda$ forces, the contribution of $\Lambda\Lambda$
interactions to the EoS is determined by $\lambda_0$, $\lambda_1$ and the density of $\Lambda$ hyperon $\rho_\Lambda$.
Second, as can be seen in Fig.~\ref{fig:frac}, fraction of $\Lambda$ hyperons at the maximum mass density
(denoted by circles in the figure) is similar to each other among different $NN+N\Lambda$ interactions.
With similar $\Lambda$ hyperon fraction at the center of NS where TOV equation is started to be integrated,
the contribution of $\Lambda\Lambda$ interaction is dominantly controlled by $\lambda_0$ and
$\lambda_1$ values. 
Therefore, a $\Lambda\Lambda$ interaction model gives similar contribution to the NS maximum mass
with little dependence on the $NN$ and $N\Lambda$ interactions.
Some errors are inevitable due to the inconsistent combination of $NN$, $N\Lambda$ and $\Lambda\Lambda$ 
interactions, but it is unlikely to affect or change the conclusion of this work significantly.

Another limitation of the current work arises from the omission of $\Sigma$ and $\Xi$ hyperons.
Some recent works \cite{aa2012, schulze2011, ohnishi2014} consider the role of these hyperons, 
and the results turn out to be sensitive to the hyperon-nucleon potentials.
Due to the lack of sufficient experimental data, the contributions of $\Sigma$ and
$\Xi$ hyperons are more uncertain than the those of $\Lambda$ hyperon.
%
Though our investigation is not complete at this moment in the aspect of the 
self-consistent application of models and the full account of exotic degrees of freedom, 
we expect that these two effects are not large enough to alter the conclusion of this work. 
However, it is still worth continuing to investigate them in the future studies.

\begin{acknowledgments}
YL was supported by the Rare Isotope Science Project of Institute for Basic Science funded by Ministry of Science, 
ICT and Future Planning and National Research Foundation of Korea (2013M7A1A1075766). 
CHH is grateful to the Institute for Basic Science, where part of the work was completed. 
Work of CHH was supported by Basic Science Research Program through the National
Research Foundation of Korea (NRF) funded by the Ministry of Education (NRF-2014R1A1A2054096).
KK was supported by Basic Science Research Program through the National Research Foundation of Korea 
(NRF) funded by the Ministry of Science, ICT, and Future Planning (NRF-2014M1A7A1A03029872). 
CHL was supported by the National Research Foundation of Korea (NRF) grant funded by the Korea government 
(MSIP) (No. NRF-2015R1A2A2A01004238).
\end{acknowledgments}


\begin{references}

\bibitem{lattimer2012}
J.~M. Lattimer, Ann. Rev. Nucl. Part. Sci. {\bf 62}, 485 (2012).

\bibitem{1614}
P. Demorest, T. Pennucci, S. Ransom, M. Roberts, and J.~W.~T. Hessels,
Nature {\bf 467}, 1081 (2010).

\bibitem{0348}
J. Antoniadis {\it et al.},
Science {\bf 340}, 448 (2013).

\bibitem{lattimer2007}
J.~M. Lattimer and M. Prakash,
Phys. Rept. {\bf 442}, 109 (2007).

\bibitem{weiss2012}
S. Weissenborn, D. Chatterjee, and J. Schaffner-Bielich,
Phys. Rev. C {\bf 85}, 065802 (2012).

\bibitem{weiss2013}
S. Weissenborn, D. Chatterjee, and J. Schaffner-Bielich,
Nucl. Phys. A {\bf 914}, 421 (2013).

\bibitem{hyun2007}
C.~H. Hyun, Prog. Theor. Phys. Suppl. {\bf 168}, 627 (2007).

\bibitem{bed2005}
I. Bednarek, and R. Manka,
J. Phys. G: Nucl. Part. Phys. {\bf 31}, 1009 (2005).

\bibitem{ryu2007}
C.~Y. Ryu, C.~H. Hyun, S.~W. Hong, and B.~T. Kim,
Phys. Rev. C {\bf 75}, 055804 (2007).

\bibitem{ryu2009}
C.~Y. Ryu, C.~H. Hyun, and S.~W. Hong,
J. Korean Phys. Soc. {\bf 54}, 1448 (2009). 

\bibitem{aa2012}
I. Bednarek, P. Haensel, J.~L. Zdunik, M. Bejger, and R. Manka,
Astron.Astrophys. {\bf 543}, A157 (2012).

\bibitem{lim2014}
Y. Lim, K. Kwak, C.~H. Hyun, and C.-H. Lee,
Phys. Rev. C {\bf 89}, 055804 (2014).

\bibitem{steiner2010}
A.~W. Steiner, J.~M. Lattimer, and E.~F. Brown,
ApJ {\bf 722}, 33 (2010).

\bibitem{rayet1981}
M. Rayet, Nucl. Phys. A {\bf 368}, 381 (1981).

\bibitem{lans1997}
D.~E. Lanskoy, and Y. Yamamoto,
Phys. Rev. C {\bf 55}, 2330 (1997).

\bibitem{ybz1988}
Y. Yamamoto, H. Bando, and J. Zofka, Prog. Theor. Phys. {\bf 80}, 757 (1988).

\bibitem{fer1989}
F. Fernandez, T. Lopez-Arias, and C. Prieto, Z. Phys. A {\bf 334}, 349 (1989).

\bibitem{gul2012}
N. Guleria, S.~K. Dhiman, and R. Shyam, Nucl. Phys. A {\bf 886}, 71 (2012).

\bibitem{lans1998}
D.~E. Lanskoy, Phys. Rev. C {\bf 58}, 3351 (1998).

\bibitem{minato2011}
F. Minato, and S. Chiba, Nucl. Phys. A {\bf 856}, 55 (2011).

\bibitem{gul2014}
N. Guleria, S.~K. Dhiman, and R. Shyam,
Int. J. Mod. Phys. E {\bf 23}, 1450026 (2014).

\bibitem{naka2010}
K. Nakazawa {\it et al.}, Nucl. Phys. A {\bf 835}, 207 (2010).

\bibitem{dany1963}
M. Danysz {\it et al.}, Nucl. Phys. {\bf 49}, 121 (1963).

\bibitem{hiyama2002}
E. Hiyama, M. Kamimura, T. Motoba, T. Yamada, and Y. Yamamoto,
Phys. Rev. C {\bf 66}, 024007 (2002).

\bibitem{iwata2012}
Y. Iwata, and J.~A. Maruhn,
arXiv:1211.2355 [nuch-th].

\bibitem{mi2010}
A.-J. Mi, and W. You,
Commun. Theor. Phys. {\bf 53}, 133 (2010).

\bibitem{bs2015}
P.~F. Bedaque, and A.~W. Steiner,
Phys. Rev. C {\bf 92}, 025803 (2015).

\bibitem{mornas2005}
L. Mornas, Eur. Phys. J. A {\bf 24}, 293 (2005).

\bibitem{schulze2011}
H.-J. Schulze, and T. Rijken,
Phys. Rev. C {\bf 84}, 035801 (2011).

\bibitem{ohnishi2014}
K. Tsubakihara, A. Ohnishi, and T. Harada,
arXiv:1402.0979 [nucl-th].

\end{references}
\end{document}